\documentclass[12pt]{article}
\usepackage{latexsym, longtable, graphicx}
\usepackage{amssymb, amsmath, xspace, lscape,  latexsym}
\usepackage{epstopdf}

\renewcommand{\le}{\leqslant}

\def\beq#1#2\eeq{%
        \begin{equation}%
        \label{#1}%
            #2%
        \end{equation}%
    }

\newcommand{\mref}[1]{(\ref{#1})}

\newcommand{\p}{\partial}

\renewcommand{\a}{\alpha}
\renewcommand{\b}{\beta}
\newcommand{\g}{\gamma}

\newcommand{\C}{\mathbb C}

\renewcommand{\tilde}{\widetilde}

\def\btheor#1\etheor{%
        \begin{theor}%
            #1%
        \end{theor}%
    }

    \def\bsled#1\esled{%
        \begin{sled}%
            #1%
        \end{sled}%
    }

\newtheorem{theorem}{Theorem}

\newtheorem{lemma}{Lemma}
\newtheorem{prop}{Proposition}
\newtheorem{cor}{Corollary}

\vspace{4ex}

\def\hm#1{#1\nobreak\discretionary{}{\hbox{\m@th$#1$}}{}}
\def\mi#1{\discretionary{\hbox{\m@th$#1$}}{\hbox{\m@th$#1$}}{}}

\begin{document}

\begin{center}

{\bf\Large On the geometry of $\vee$-systems}

\vspace{1cm}

{\bf  M. Feigin$^*$, A.P. Veselov$^\dagger$ }
\end{center}

\vspace{1cm} \noindent$^*$ Department of Mathematics, University of
Glasgow, University Gardens, Glasgow G12 8QW, UK. Email:
m.feigin@maths.gla.ac.uk

\noindent$^\dagger$ Department of Mathematical Sciences,
Loughborough University, Loughborough, Leicestershire, LE11 3TU, UK;
Landau Institute for Theoretical Physics, Kosygina 2, Moscow,
117940, Russia. Email: A.P.Veselov@lboro.ac.uk

\vspace{1cm}
\rightline{\em Dedicated to Sergey Petrovich Novikov on his 70th birthday}
\bigskip

\begin{abstract}
We consider a complex version of the $\vee$-systems, which appeared in the theory of the WDVV equation.  
We show that the class of these systems is closed under the natural operations of restriction and taking the subsystems and study a special class of the $\vee$-systems
related to generalized root systems and basic classical Lie superalgebras.
\end{abstract}

\section{Introduction}

The main object of our study is the special collections of vectors
in a linear space, which are called $\vee$-systems. They were introduced in \cite{V1,V2} in relation with a certain class of special solutions of the generalized Witten-Dijkgraaf-Verlinde-Verlinde (WDVV) equations, playing an important role in 2D topological  field theory and $N=2$ SUSY Yang-Mills theory \cite{D1,MMM}. A geometric theory of the WDVV equation was developed by Dubrovin, who introduced a fundamental notion of Frobenius manifold \cite{D1,D2}.

The definition of the $\vee$-systems is as follows. Let for beginning $V$ be a real vector space and  $\mathcal{A}\subset V^*$ be a
finite set of vectors in the dual space $V^*$ (covectors) spanning $V^*$. To such a set one can
associate the following {\it canonical form} $G_{\mathcal A}$ on
$V$:
$$G_{\mathcal A}(x,y)=\sum_{\a\in\mathcal{A}}\a(x)\a(y), $$
where $x,y\in V$. This is a non-degenerate scalar product, which
establishes the isomorphism
$$
\varphi_{\mathcal A}: V \rightarrow V^*.
$$
The inverse $\varphi_{\mathcal A}^{-1}(\alpha)$ we denote as
$\a^\vee$.
The system $\mathcal{A}$ is called $\vee$-{\it system} if the following relations (called $\vee$-{\it conditions})
$$
\sum\limits_{\beta \in \Pi \cap \mathcal {A}}
\beta(\alpha^\vee)\beta^\vee=\lambda \alpha^{\vee}
$$
are satisfied for any $\alpha \in \mathcal{A}$ and any two-dimensional plane $\Pi \subset V^*$ containing $\alpha$ and some $\lambda$, which may depend on $\Pi$ and $\alpha.$ For more geometric definition and the relation to WDVV equation see \cite{V1,V2} and the next section.

One can show \cite{V1} that all Coxeter root systems as well as their deformed versions appeared in the theory of quantum Calogero-Moser systems satisfy these conditions  (see also \cite{CV, V3}, where the relation between $\vee$-systems and Calogero-Moser theory was clarified).

In this paper we study the geometric properties of $\vee$-systems in more detail.
In particular we show that a subsystem of the $\vee$-system is also a $\vee$-system, the result which we announced in \cite{F}. This may be not surprising but not obvious from the definition.

A surprising fact is that the restriction of a $\vee$-system
$\mathcal A$ to the  subspace defined by a subset $\mathcal B
\subset \mathcal A$ is also a $\vee$-system (see \cite{FV}). This is
clearly not true for the Coxeter root systems. In fact,
$\vee$-systems can be considered as an extension of the class of
Coxeter systems, which has this property.

We show that all these properties (under some mild additional assumptions) are true also for a natural complex version of the $\vee$-systems, which we discuss in the next section. The consideration of the complex $\vee$-systems  was partly motivated by the link with the theory of Lie superalgebras developed in \cite{SV1}. We study the new examples of the $\vee$-systems coming from this theory in relation with the restrictions of Coxeter root systems investigated in \cite{FV}.

In the last section we discuss complex Euclidean $\vee$-systems, which is an extension of the class of  $\vee$-systems, when the canonical form is allowed to be degenerate. The root systems of some basic classical Lie superalgebras give important examples of such systems.

We finish with the list of all known $\vee$-systems in dimension 3.

\section{Complex $\vee$-systems and WDVV equation}

Let now $V$ be a complex vector space and $\mathcal{A} \subset V^*$ be a finite set of covectors. We will
assume that the bilinear form on $V$ \beq{GC} G_{\mathcal
A}(x,y)=\sum_{\a\in\mathcal{A}}\a(x)\a(y) \eeq is
non-degenerate. In the real case this would simply mean that the elements of $\mathcal A$ span $V^*$,
in the complex case our assumption is stronger. This form then
establishes the isomorphism
$$
\varphi_{\mathcal A}: V \rightarrow V^*.
$$
We denote $\a^\vee = \varphi_{\mathcal A}^{-1}(\alpha)$ and say in full analogy with the real case \cite{V1} that
$\mathcal{A}$ is a $\vee$-{\it system} if for any $\a\in \mathcal A$
and for any two-dimensional plane $\pi$ containing $\alpha$ the
following $\vee$-condition holds \beq{acheckComplex} \sum_{\b\in\pi\cap
\mathcal{A}} \b(\a^\vee) \b^\vee = \lambda \a^\vee \eeq for some constant
$\lambda=\lambda(\a,\pi)$. Equivalently one can say that
either subsystem $\Pi=\pi \cap \mathcal{A}$ is reducible in the sense that it consists of two orthogonal
subsystems or the
following forms are proportional:
$$
G_{\Pi}|_{\pi^\vee \times V} \sim G_\mathcal{A}|_{\pi^\vee \times
V},
$$
where
\beq{gpi}
G_{\Pi}(x,y)=\sum_{\b\in\Pi \cap \mathcal A}\b(x)\b(y).
\eeq

Originally $\vee$-systems in $\mathbb{R}^n$ appeared as geometric
reformulation of the Witten-Dijkgraaf-Verlinde-Verlinde equations
for the prepotential \beq{prep} F=\sum_{\a\in \mathcal A} \a(x)^2
\log \a(x)^2, \eeq but the proof \cite{V1} was using some geometry of the real plane.
We will show now that one can avoid this and that similar
interpretation holds in the complex case as well.

Recall first that the
(generalized) WDVV equations have the form \beq{wdvv} F_i F_j^{-1}
F_k = F_k F_j^{-1} F_i \eeq
 for any
$i,j,k=1,\ldots,n,$ where $F_i$ is the matrix of third derivatives
$(F_i)_{ab}=\p^3F/{\p x^i \p x^a \p x^b}$ (see \cite{MMM}). As it
was explained in \cite{MMM2} the system \mref{wdvv} is equivalent to
the system \beq{wdvv2} F_i G^{-1} F_j = F_j G^{-1} F_i \eeq
 for any
$i,j=1,\ldots,n,$ where $G$ is any non-degenerate linear combination
$G=\sum_{i=1}^n \eta^i(x)F_i$. For instance, for the prepotential
\mref{prep} choosing $\eta^i(x)=\frac14 x^i$ one arrives at
$x$-independent form \beq{wdvvGa} G=G_{\mathcal A} = \sum_{\a\in
\mathcal A} \a \otimes \a. \eeq
 The following lemma can be proved exactly like
in the real case \cite{V2}.
\begin{lemma}\label{lem1}
The WDVV equations \mref{wdvv2}, \mref{wdvvGa} for the prepotential
\mref{prep} are equivalent to the identities \beq{pi}
 \sum_{\b\in
\pi \cap \mathcal A} G_{\mathcal A}(\a^\vee, \b^\vee)
\left(\a(a)\b(b)-\a(b)\b(a)\right)= 0
\eeq
 for any $\a \in \mathcal A$, any
2-plane $\pi$ containing $ \a$ and arbitrary $a,b \in V$.
\end{lemma}

Indeed, a direct substitution of the form  \mref{prep} into the WDVV equation  \mref{wdvv2} gives for arbitrary $a,b$
\beq{sc}
\sum\limits_{\alpha \ne \beta, \alpha,\beta \in \mathcal{A}}
\frac{G_\mathcal{A} (\alpha^\vee, \beta^\vee)B_{\alpha,\beta}(a,b)}
{(\alpha,x)(\beta,x)}\alpha\wedge \beta \equiv 0,
\eeq
where
$
\alpha\wedge \beta=\alpha\otimes \beta-\beta \otimes \alpha
$
and
$
B_{\alpha,\beta}(a,b)=\alpha(a)\beta(b)-\alpha(b)\beta(a).
$
It is easy to see that these relations can be rewritten as
$$
\sum\limits_{\beta \neq \alpha, \beta \in \pi \cap \mathcal{A}}
\frac{G_\mathcal{A} (\alpha^\vee, \beta^\vee) B_{\alpha,\beta}(a,b)}
{(\beta,x)}\alpha \wedge \beta |_{(\alpha,x)=0}\equiv 0
$$
for any $\alpha \in \mathcal{A}$ and any two-dimensional plane $\pi$ containing
$\alpha$, which are equivalent to \mref{pi}.

We are now ready to show the equivalence of $\vee$-conditions and the WDVV
equations for the prepotential \mref{prep} in the complex case.

\begin{theorem}\label{theor1}
 The prepotential \mref{prep} satisfies the WDVV
equations \mref{wdvv2}, \mref{wdvvGa} if and only if $\mathcal A$ is
a $\vee$-system.
\end{theorem}
{\bf Proof.} By Lemma \ref{lem1} the WDVV equations are equivalent
to the identities \beq{checkid}
 \sum_{\b\in
\Pi} \b(\a^\vee) \left(\a(a)\b(b)-\a(b)\b(a)\right)=0 \eeq
 for any $\a \in \mathcal A$, for any
two-dimensional subsystem $\Pi= \pi \cap \mathcal A \ni \a$, for any
$a,b \in V$. Relation \mref{checkid} can be rewritten as
$$
\a(a)G_\Pi(\a^\vee,b)=\a(b)G_\Pi(\a^\vee,a).
$$
Therefore the ratio $G_\Pi(\a^\vee,a)/\a(a)$ does not depend on
vector $a \in V$ and we can further rewrite \mref{checkid} as
\beq{checkidsimple} G_\Pi(\a^\vee,a)=\lambda \a(a) = \lambda
G_{\mathcal A}(\a^\vee,a), \eeq where
$\lambda=\lambda(\a,\Pi)=const$. Consider now a linear operator
$A_\Pi$ defined by the pair of these bilinear forms:
$$
G_\Pi(\a^\vee,a)=G_{\mathcal A} (A_\Pi \a^\vee, a)
$$
for any $a \in V$. The property \mref{checkidsimple} states that for any
$\a \in \Pi$ the vector $\a^\vee$ is an eigenvector of the operator $A_\Pi$. In case
when $\Pi$ contains at least three pairwise non-collinear covectors we
conclude that $A_\Pi|_{\pi^\vee}$ is scalar, so
$$
G_\Pi|_{\pi^\vee \times V} = \lambda G_{\mathcal A}|_{\pi^\vee
\times V}
$$
which is a $\vee$-condition. If $\Pi$ contains only two non-collinear covectors $\a, \b$ we choose $a\in V$ so that
$\a(a)=0$, $\b(a)\ne 0$. Then \mref{checkidsimple} states
$\b(\a^\vee)=0$ hence $\Pi$ is reducible. Theorem is proven.

\section{Subsystems and restrictions of $\vee$-systems}\label{firstsec}

Let  $\mathcal A \subset V$ be a $\vee$-system in a real or complex vector space $V.$
The subset $\mathcal B \subset \mathcal A$ is called a {\it
subsystem} if $\mathcal{B}=\mathcal{A}\cap W$ for some vector subspace
$W$. We will assume that the corresponding space $W$ is
spanned by $\mathcal{B}$. The {\it dimension of $\mathcal{B}$} is by
definition the dimension of the subspace $W$. Subsystem
$\mathcal{B}$ is called {\it reducible} if
$\mathcal{B}=\mathcal{B}_1 \sqcup \mathcal{B}_2$ is a union of two
non-empty subsystems orthogonal with respect to the canonical form
on $V\cong V^*$.

Consider the following bilinear form on $V$ \beq{gp}
G_{\mathcal B}(x,y)=\sum_{\b\in\mathcal{B}}\b(x)\b(y) \eeq
associated with subsystem $\mathcal{B}$.
The subsystem $\mathcal B$ is called
{\it isotropic} if the restriction $G_{\mathcal B}|_{W^\vee}$ of the form $G_{\mathcal B}$ onto the subspace $W^\vee \subset V$
is degenerate and {\it non-isotropic} otherwise.

\begin{theorem}\label{t0}
Any non-isotropic subsystem $\mathcal B$ of a $\vee$-system
$\mathcal A$ is also a $\vee$-system.
\end{theorem}
{\bf Proof.} Consider the operator
$$
A_W=\sum_{\beta \in \mathcal B} \beta \otimes \beta^\vee : W^\vee
\to W^\vee.
$$
For any $\a \in \mathcal B$ the vector $\alpha^\vee$ is an
eigenvector for $A_W$. Indeed, this follows from summing up the
$\vee$-conditions \mref{acheckComplex} over the 2-planes $\pi$ such
that $\alpha \in \pi \subset W$. Since $\alpha^\vee$ span $W^\vee$ we have the eigenspaces
decomposition
$$
W^\vee = U_1 \oplus U_2 \oplus \ldots \oplus U_k
$$
where $A_W|_{U_i}= \lambda_i I$ are scalar operators. Let vector $u \in U_i$
and $v \in V$. Then we have
$$
\lambda_i G_{\mathcal A}(u,v) = G_{\mathcal A}(A_W u, v) =
G_{\mathcal B}(u,v).
$$
Therefore \beq{red} G_{\mathcal B}|_{U_i \times U_j} = G_{\mathcal
A}|_{U_i \times U_j} =0 \eeq
 for $i\ne j$, and
\beq{red2} G_{\mathcal B}|_{U_i \times V} = \lambda_i G_{\mathcal
A}|_{U_i \times V}. \eeq
 Now we are ready to verify the
$\vee$-conditions for the subsystem $\mathcal B$ of covectors on
$W^\vee$. By assumption of non-isotropicity  the form $G_{\mathcal
B}|_{W^\vee}$ is non-degenerate. This implies that all $\lambda_i
\ne 0$. Consider now a 2-plane $\pi \subset  W^\vee$. If $\pi$
nontrivially intersects two summands $U_i$ and $U_j$ then property
\mref{red} implies reducibility of $\pi$ with respect to
$G_{\mathcal B}$. If $\pi \subset U_i$ for some $i$ then the
$\vee$-condition \mref{acheckComplex} for $\mathcal A$ implies the
$\vee$-condition for $\mathcal B$ as taking $\vee$ with respect to
$G_{\mathcal A}$ and $G_{\mathcal B}$ differs by constant multiplier
$\lambda_i$ on $U_i$. Theorem is proven.

Same arguments show that the definition of the $\vee$-systems can be
reformulated in a more natural way.

Recall that $\vee$-systems can be defined as finite sets
$\mathcal{A} \subset V^*$ such that for any two-dimensional
subsystem $\mathcal{B}=\mathcal{A}\cap\pi$ either the restrictions
of bilinear forms $G_\mathcal{B}$ and $G_\mathcal{A}$ to
$\pi^\vee\times V$ are proportional or subsystem $\mathcal{B}$ is
reducible (see \cite{V1} and previous section).

We claim that in this definition the restriction on the dimension
of the subsystem $\mathcal{B}$ can be omitted.

\begin{theorem}\label{t1}
For {\bf any} subsystem $\mathcal{B}=\mathcal{A}\cap W$ of a
$\vee$-system $\mathcal{A}$ either $G_\mathcal{B}|_{W^\vee\times V}$
and $G_\mathcal{A}|_{W^\vee \times V}$ are proportional or
$\mathcal{B}$ is reducible.
\end{theorem}
{\bf Proof.} As we established in the proof of Theorem \ref{t0} the
space $W^\vee$ can be decomposed as
$$
W^\vee = U_1 \oplus U_2 \oplus \ldots \oplus U_k
$$
so that relations \mref{red}, \mref{red2} hold. In the case $k>1$
the property \mref{red} implies reducibility of the subsystem
$\mathcal{B}$. In the case $W^\vee=U_1$ the property \mref{red2}
states required proportionality of restricted bilinear forms.

\begin{cor}\label{cor}
The $\vee$-systems can be defined as the finite sets $\mathcal A
\subset V^*$ with non-degenerate form $G_\mathcal{A}$ such that for
any subsystem $\mathcal{B}=\mathcal{A}\cap W$ of a $\vee$-system
$\mathcal{A}$ either $G_\mathcal{B}|_{W^\vee \times V}$ and
$G_\mathcal{A}|_{W^\vee \times V}$ are proportional or $\mathcal{B}$
is reducible.
\end{cor}

Let us consider now the restriction operation for the $\vee$-systems.  For any subsystem $\mathcal{B}\subset \mathcal A$
consider the corresponding subspace $W_\mathcal{B}\subset V$ defined
as the intersection of hyperplanes
$$
\b(x)=0,\,\, \b \in \mathcal{B}.
$$
Let the set $\pi_\mathcal{B}(\mathcal{A})$ consist of the
restrictions of covectors $\a\in \mathcal{A}$ on $W_\mathcal{B}$.

Similar to the real case \cite{FV} we claim that the class of the $\vee$-systems is closed under this operation.

\begin{theorem}\label{theor4} Assume that the restriction $G_{\mathcal
A}|_{W_\mathcal{B}}$ is non-degenerate. Then the restriction
$\pi_\mathcal{B}(\mathcal{A})$ of a $\vee$-system $\mathcal{A}$ is
also a $\vee$-system.
\end{theorem}

The proof is parallel to the real case \cite{FV}.
It uses the notion of {\it logarithmic Frobenius structure} \cite{D1, FV} and is based on the following two lemmas.

Let $M_{{\mathcal{A}}} = V \setminus \cup_{\alpha \in {\mathcal{A}}}
\Pi_{\alpha}$ be the complement to the union of all hyperplanes
$\Pi_{\alpha}: \alpha(x)=0$ and similarly $M_{{\mathcal{B}}} =
W_\mathcal{B} \setminus \cup_{\alpha \in
{{\mathcal{A}}\setminus{\mathcal{B}}}} \Pi_{\alpha}.$ Consider the
following multiplication on the tangent space $T_xM_{{\mathcal{A}}}$
on $M_{{\mathcal{A}}}$: \beq{mult} u * v = \sum_{\a\in \mathcal A}
\frac{\a(u) \a(v)}{\a(x)} {\a^\vee}. \eeq

\begin{lemma}\label{WDVVassoc}
The multiplication  (\ref{mult}) on the tangent bundle
$T_*M_{{\mathcal{A}}}$ is associative iff
$$
F_{\mathcal A} = \sum_{\a\in \mathcal{A}} \a(x)^2 \log \a(x)^2,
\quad x \in M_{{\mathcal{B}}}
$$
satisfies the WDVV equation \mref{wdvv2}, \mref{wdvvGa}.

\end{lemma}

Consider now a point $x_0 \in M_{{\mathcal{B}}}$ and two tangent vectors $u,v$ at $x_0$ to $M_{{\mathcal{B}}}.$
We extend vectors $u$ and $v$ to two local analytic vector fields
$u(x), v(x)$ in the neighbourhood of $x_0 \in V$, which are tangent
to the subspace $W_\mathcal{B}$.

\begin{lemma}\label{lem0}
The product $u(x) * v(x)$ has a limit when $x$ tends to $x_0$ given
by \beq{mult2} u * v = \sum_{\a \in \mathcal{A}\backslash
\mathcal{B}} \frac{\a(u) \a(v)}{\a(x)} {\a^\vee}. \eeq The limit is
determined by $u$ and $v$ only and is tangent to the subspace
$W_\mathcal{B}.$
\end{lemma}

Using the orthogonal decomposition $V=W_\mathcal{B} \oplus
{W_\mathcal{B}}^\perp$ one can rewrite \mref{mult2} as \beq{mult25}
u
* v = \sum_{\a \in \mathcal{A}\backslash \mathcal{B}} \frac{\a(u)
\a(v)}{\a(x)} {\tilde {\a^\vee}}, \eeq where $\tilde {\a^\vee}$ is
orthogonal projection of the vector $\a^\vee$ to $W_\mathcal{B}$.
The vector $\tilde {\a^\vee} \in W_\mathcal{B}$ can be shown to be
dual to the  covector $\pi_\mathcal{B}(\a)$ under the canonical form
restricted to $W_\mathcal{B}$. Therefore the associative
multiplication \mref{mult25} is determined by the prepotential
\beq{potr} F_{\mathcal B} = \sum_{\a\in \mathcal{A} \backslash
\mathcal{B}} \a(x)^2 \log \a(x)^2, \quad x \in M_{{\mathcal{B}}}.
\eeq By Lemma \ref{WDVVassoc} prepotential $F_{\mathcal B}$
satisfies the WDVV equation and Theorem \ref{theor4} follows from
Theorem~\ref{theor1}.

\section{$\vee$-systems and generalized root systems}

Generalized root systems were introduced by
Serganova in relation to basic classical Lie superalgebras
\cite{Sergan}. They are defined as follows.

Let $V$ be a finite dimensional complex vector space with a
non-degenerate bilinear form $< , >$.
The finite set $R\subset V\setminus\{0\}$ is
called a {\it generalized root system} if the following conditions
are fulfilled :

1) $R$ spans $V$ and $R=-R$ ;

2) if $\alpha,\beta\in R$ and  $<\alpha ,\alpha >\ne 0$ then
$\frac{2<\alpha ,\beta >}{<\alpha ,\alpha >}\in {\bf Z}$ and
$s_{\alpha}(\beta)=\beta -\frac{2<\alpha ,\beta >}{<\alpha ,\alpha
>}\alpha\in R$;

3) if $\alpha\in R$ and $<\alpha ,\alpha >=0$ then for any
$\beta\in R$ such that $<\alpha ,\beta >\ne 0$ at least one of the
vectors $\beta + \alpha$ or $\beta - \alpha$ belongs to $R$.

Any generalized root system has a partial symmetry described by the
finite group $W_0$ generated by reflections with respect to the
non-isotropic roots.

Serganova classified all irreducible generalised root systems. The
list consists of classical series $ A(n,m)$ and $BC(n,m)$ and three
exceptional cases $G(1,2) $, $AB(1,3)$ and $D(2,1,\lambda),$ which
essentially coincides with the list of basic classical Lie
superalgebras.

In the paper \cite{SV1} Sergeev and one of the authors introduced a
class of {\it admissible deformations} of generalized root systems,
when the bilinear form $<,>$ is deformed and the roots $\alpha \in
R$ acquire some multiplicities $m_{\alpha}.$ They satisfy the
following 3 conditions:

1) the deformed form $B$ and the multiplicities are $W_{0}$-invariant;

2) all  isotropic roots have multiplicity 1;

3) the function $\psi_{0}=\prod_{\alpha\in R_{+}} \sin^{-m_{\alpha}}(\alpha,x)$ is a (formal) eigenfunction of
the  Schr\"odinger operator
\begin{equation}
\label{dCMS}
L=-\Delta + \sum_{\alpha\in R_{+}}\frac{m_{\alpha}(m_{\alpha}+2m_{2\alpha}+1)(\alpha,\alpha)}{\sin^2(\alpha,x)},
\end{equation}
where the brackets ( , ) and the Laplacian $\Delta$ correspond to
the deformed bilinear form $B,$ which is assumed to be
non-degenerate.

All admissible deformations of the generalized root systems were described explicitly in \cite{SV1}.
They depend on several parameters, one of which (denoted $k$ in \cite{SV1}) describes the deformation of the bilinear form, so that the case $k=-1$ corresponds to the original generalized root system.

The following theorem follows from the results of \cite{V3, SV1}.

\begin{theorem}\label{sup}
For any admissible deformation  $(R, B, m)$ of a generalized root
system $R$ the set $\mathcal A = \{\sqrt{m_\alpha}\alpha,\,\, \alpha
\in R\}$ is a $\vee$-system whenever the canonical form
 $$
G_{\mathcal A}(u,v)=\sum_{\a \in \mathcal A} m_\a \a(u) \a(v)
 $$
 is non-degenerate.
 In particular, for any basic classical Lie superalgebra $\mathfrak{g}$ with non-degenerate Killing form the set $\mathcal A_{\mathfrak{g}},$ consisting of the even roots of $\mathfrak{g}$ and the odd roots multiplied by $i = \sqrt{-1},$ is a $\vee$-system.
 \end{theorem}

The canonical form (\ref{GC}) for the system  $\mathcal A_{\mathfrak{g}}$ coincides with the Killing form of the corresponding Lie superalgebra $\mathfrak{g}.$
Note that in contrast to the simple Lie algebra case the Killing form of basic classical Lie superalgebra could be zero, which is the case only for the Lie superalgebras of type $A(n,n)$, $D(n+1,n)$ and  $D(2,1,\lambda).$

The $\vee$-systems corresponding to the classical generalized root
systems $A(n,m)$ and $BC(n,m)$ are particular cases of the
following multiparameter families of the $\vee$-systems $A_n(c)$ and $B_n(\gamma;
c)$ (appeared in \cite{CV}, see also \cite{FV}): the $\vee$-system $A_n(c)$ consists of the covectors
$$
\sqrt{c_i c_j} (e_i-e_j),\,\,\,  1\le i <j \le n+1,
$$
and the $\vee$-system $B_n(\gamma; c)$ consists of
$$
\sqrt{c_i c_j} (e_i\pm e_j),\, 1\le i <j \le n; \quad
\sqrt{2c_i(c_i+\gamma)}e_i, \, 1\le i \le n.
$$

We will also need Coxeter root system $B_n(t)$ consisting of the
covectors
$$
e_i \pm e_j, \, 1\le i<j \le n; \quad \quad  t e_i, \, 1 \le i \le
n.
$$
Like in \cite{FV}, we will denote by $(\mathcal A, \mathcal B)$ the
restrictions of a $\vee$-system $\mathcal A$ along the $\vee$-system
$\mathcal B$. We will use the subindexes $i$ as ${(\mathcal A,
\mathcal B)}_i$ if there are a few embeddings of a $\vee$-system
$\mathcal B$ into $\mathcal A$ leading to non-equivalent
restrictions.

Now we are going to study in more detail the $\vee$-systems coming from the exceptional Lie superalgebras.

The family of $\vee$-systems $AB_4(t)$ corresponding to the
exceptional four-dimensional generalized root system  $AB(1,3)$ is
analyzed in \cite{FV} (see section 6) in relation to the
restrictions of Coxeter $\vee$-systems. We only mention here that
this family has two different non-Coxeter three-dimensional
restrictions $(AB_4(t), A_1)_1$, $(AB_4(t), A_1)_2,$ consisting of
the following covectors
$$
(AB_4(t),A_1)_1:  \sqrt{2(2t^2+1)} e_1, \, 2 \sqrt{2(t^2+1)} e_2, t
\sqrt{\frac{2(2t^2-1)}{t^2+1}}e_3, $$ $$ \sqrt{2}(e_1\pm e_2), \,
t\sqrt{2}(e_1 \pm e_3), \, t(e_1 \pm 2 e_2 \pm e_3);
$$
$$
(AB_4(t),A_1)_2: e_1+e_2, e_1+e_3, e_2+e_3, \sqrt{2}e_1,
\sqrt{2}e_2, \sqrt{2}e_3,
\frac{t\sqrt{2}}{\sqrt{t^2+1}}(e_1+e_2+e_3),
$$
$$
\frac1{\sqrt{4t^2+1}}(e_1-e_2), \frac1{\sqrt{4t^2+1}}(e_1-e_3),
\frac1{\sqrt{4t^2+1}}(e_2-e_3).
$$

In the case of the exceptional generalized root system
$G(1,2)$ the corresponding family of $\vee$-systems, which we denote $G_3(t)$,
 consists of the following  covectors (see \cite{SV1}):
\beq{g3t} \sqrt{2t+1}e_1, \sqrt{2t+1}e_2, \sqrt{2t+1}(e_1+e_2),
\sqrt{\frac{2t-1}3}(e_1-e_2), \sqrt{\frac{2t-1}3}(2 e_1+e_2), \eeq
$$
\sqrt{\frac{2t-1}3}(e_1+ 2 e_2), \,\, \sqrt{\frac{3}{t}} e_3, \,\,
e_1 \pm e_3,\,\,  e_2 \pm e_3, \,\, e_1+e_2 \pm e_3.
$$

\begin{theorem}\label{teorg}
The set of covectors $G_3(t)$ with
$t\ne 0, -\frac12$ is a $\vee$-system, which is equivalent to a
restriction of a Coxeter root system if and only if $t=1$ or $t=3/4$
or $t=1/2$. The corresponding Coxeter restrictions are
$(E_7,A_2^2)$, $(E_8,A_5)$ and $(E_6, A_1^3)$ respectively.
\end{theorem}
{\bf Proof.}  One can check that the corresponding canonical form
\mref{GC} is degenerate if and only if $t=-\frac12.$ Together with
Theorem \ref{sup} this implies the first claim.

To establish the equivalences with the restrictions of Coxeter root
systems note that if $t \ne \pm 1/2$ the system contains 13 pairwise
non-parallel covectors. All the Coxeter restrictions are given
explicitly in \cite{FV}. In particular, it is shown that there is a
one-parameter family $F_3(\lambda)$ of $\vee$-systems  with 13
covectors in dimension 3. This family is a restriction of the
Coxeter $\vee$-system $F_4(\lambda)$ and contains Coxeter
restrictions $(E_7, A_1\times A_3)_1$, $(E_7, A_1^4)$, $(E_8, D_5)$,
$(E_8, A_1\times D_4)$ (see \cite{FV}). Any $\vee$-system from this
family does not have a two-dimensional plane, containing more than 4
covectors. Since the system $G_3(t)$ has 6 covectors in the plane
$\langle e_1,e_2\rangle$, it is not equivalent to those Coxeter
restrictions.

The three-dimensional Coxeter restrictions not belonging to the
$F_3(\lambda)$ family and containing 13 covectors are $(E_7, A_2^2)$
, $(E_8,A_5)$ and $(E_7, A_1^2 \times A_2)$. To compare $G_3(t)$
with these systems, we compare the lengths of covectors. One can
check that $G_3(t)$ has three covectors with  length squared 1/6,
three covectors  with  length squared $(2t-1)/(12 t+6)$, six
covectors  with  length squared $(t+1)/(12t+6)$ and one covector
with  length squared $1/(4t+2)$. These lengths cannot match the
lengths in the system $(E_7, A_1^2 \times A_2)$. They match the
lengths in $(E_7, A_2^2)$, $(E_8,A_5)$ if and only if $t=1$ and
$t=3/4$ respectively. It is easy to find a linear transformation
mapping $G_3(1)$ to $(E_7, A_2^2)$, and another transformation
mapping $G_3(3/4)$ to $(E_8,A_5)$.

In the remaining case $t= 1/2$ the corresponding $\vee$-system $G_3(1/2)$
consists of 10 non-parallel covectors. One can show that
it is equivalent to $(E_6, A_1^3)$.  This completes the proof.

 The $\vee$-systems corresponding to the last
(family of) exceptional generalized root systems $D(2,1,\lambda)$
consist of the following  covectors in
$\C^3$  \beq{d3lambda} e_1\pm e_2
\pm e_3, \sqrt{2(-1+t+s)} e_1, \sqrt{\frac{2(s-t+1)}{t}} e_2,
 \sqrt{\frac{2(t-s+1)}{s}} e_3, \eeq
where $t, s$ are two parameters. They are related to the projective parameters
$\lambda=(\lambda_1:\lambda_2:\lambda_3)$ as follows $$t=
\frac{\lambda_2}{\lambda_1},\, s = \frac{\lambda_3}{\lambda_1}.$$
The corresponding form (\ref{GC}) is degenerate if and only if
$t+s+1=0,$ which corresponds to the Lie superalgebra case $\lambda_1
+ \lambda_2 + \lambda_3 =0.$ We denote this family of covectors as
$D_3(t,s)$ assuming that $t+s+1 \neq 0.$

\begin{theorem}\label{teord}
The sets of covectors $D_3(t,s)$ is a two-parametric family of
$\vee$-systems. The one-parameter subfamilies $D_3(t,t)$,
$D_3(t,1)$, $D_3(1,t)$ are equivalent to the family of Coxeter
restrictions $B_3(-1; 1, 1, s)$. The one-parameter subfamilies
$D_3(t,t-1), D_3(t, -t+1), D_3(t, t+1)$ are equivalent to the family
of Coxeter restrictions $A_3(s,s,1,1)$. There are no other
intersections of the $D_3$-family with $A_3$- and $B_3$-families.
\end{theorem}
{\bf Proof.} The fact that  $D_3(t,s)$ is a $\vee$-system follows
from Theorem \ref{sup}. The following equivalences can be
established by finding appropriate linear transformations:
$$
D_3(t,t) = D_3(1/t,1) = D_3(1,1/t) = B_3(-1; 1, 1, 2t),
$$
$$
D_3(t,t-1)=A_3(t-1,t-1,1,1), \, \quad
D_3(t,-t+1)=A_3(\frac{1-t}{t},\frac{1-t}{t},1,1), \,
$$
$$
D_3(t,t+1)=A_3(t,t,1,1).
$$
To find all the intersections of $D_3$-family with $B_3$-family we
note that the corresponding $\vee$-systems from $B_3(\gamma; c_1,
c_2, c_3)$-family must have parameters (up to reordering $c_i$)
$c_1=c_2=-\gamma$ in order to consist of 7 covectors. Then it takes
the form
$$
\sqrt{-\gamma c_3}(f_1\pm f_3), \sqrt{-\gamma c_3}(f_2\pm f_3),
\gamma(f_1 \pm f_2), \sqrt{2c_3(c_3+\g)}f_3
$$
where $f_i$ are basis covectors. If there would be an equivalence with \mref{d3lambda} the covectors
\beq{ei}\sqrt{2(-1+t+s)} e_1, \sqrt{\frac{2(s-t+1)}{t}} e_2,
 \sqrt{\frac{2(t-s+1)}{s}} e_3\eeq
should be mapped to the covectors $\gamma(f_1 \pm f_2),
\sqrt{2c_3(c_3+\g)}f_3$. Then it follows that two out of three
coefficients at the covectors $e_i$ in \mref{ei} should coincide,
which leads to four possibilities $s=t$, $s=1$, $t=1$ or $s=-t-1,$
the latter is excluded.

To find all the intersections of $D_3$-family and $A_3$-family we
note that in these cases $D_3$ systems should contain 6 covectors
only. This leads to the vanishing of one of the coefficients at
$e_i$ in the formulas \mref{d3lambda}. There are three cases
$s+t=1$, $t+1=s$ or $s+1=t$. All the corresponding one-parameter
families of $\vee$-systems are presented in the formulation of the
theorem. This completes the proof.

\section{Complex Euclidean $\vee$-systems}

We have seen in the previous section that our definition of the $\vee$-systems was too rigid to include the root systems of all basic classical Lie superalgebra.  To correct this defect one can consider the following slightly more general notion, which in the real situation is equivalent to the previous case (see \cite{V2}).

Let $V$ be a {\it complex Euclidean space}, which is a complex vector space with a non-degenerate bilinear form $B$ denoted also as $ ( , ).$ We will identify $V$ with the dual space $V^*$ using this form.

Let $\mathcal A$ be a finite set of vectors in $V.$ We say that the
set $\mathcal A$ is {\it well-distributed} in $V$ if the canonical
form \beq{GEC} G_{\mathcal
A}(x,y)=\sum_{\a\in\mathcal{A}}(\a,x)(\a,y) \eeq is proportional to
the Euclidean form $B.$

We call the set $\mathcal A \subset V$ {\it complex Euclidean $\vee$-system}
if it is well-distributed in $V$ and any its two-dimensional subsystem is either reducible or well-distributed in the corresponding plane.

Note in this definition we allow the canonical form to be identically zero.
It is obvious that complex $\vee$-systems defined above can be considered as a particular case of
these systems when the canonical form \mref{GC} is non-degenerate. Indeed, in this case one can introduce a Euclidean structure on $V$ using this canonical form and all the properties in the previous definition will be satisfied.

The following version of Theorem 5 shows that there are examples of the complex Euclidean $\vee$-systems with zero canonical form.

\begin{theorem}\label{zerocf}
For any admissible deformation  $(R, B, m)$ of a generalized root
system $R$ the set $\mathcal A = \{\sqrt{m_\alpha}\alpha,\,\, \alpha
\in R\}$ is a complex Euclidean $\vee$-system. In particular, for any basic classical Lie superalgebra $\mathfrak{g}$  the set $\mathcal A_{\mathfrak{g}},$ consisting of the even roots of $\mathfrak{g}$ and the odd roots multiplied by $i = \sqrt{-1},$ is a complex Euclidean $\vee$-system.
\end{theorem}

Indeed we know that this is true for the admissible deformations with non-degenerate canonical form. Since such deformations form a dense open subset the same is true for all deformations.

We should note that complex Euclidean $\vee$-systems with zero canonical form do not determine a
logarithmic solution to WDVV equation. Indeed, the following result shows that in that case
any linear combination of the matrices $F_i$ is
degenerate, where as before
\beq{matr}
(F_i)_{jk}=\frac{\p^3F}{\p x^i \p x^j \p x^k},
\eeq
where
\beq{prepot}
F= \sum_{\a \in \mathcal A} (\a,x)^2 \log (\a,x).
\eeq

\begin{prop}
Let $\mathcal A \subset V$ be a finite collection of vectors such
that \beq{trivial}
  \sum_{\a\in \mathcal A}
(\a,u)(\a,v)\equiv 0. \eeq
Then any linear combination
$$
G=\sum_{i=1}^n \eta^i(x) F_i
$$
of matrices \mref{matr} for the corresponding prepotential \mref{prepot}, is degenerate for any $x$.
\end{prop}
{\bf Proof.} The relation \mref{trivial}  implies that for any $i,j
=1\ldots n$ \beq{3der} \sum_{k=1}^n F_{i j k} x^k = 0 \eeq and hence
$$\sum_{i,j=1}^n \eta^i (x) F_{i j k} x^k=0.$$ This means that
vector $x=(x^1,\ldots,x^n)$ belongs to the kernel of the form $G,$
which therefore is degenerate.

We should mention also that because the restriction of the complex Euclidean structure on a subspace could be degenerate the results of section 3 are true for Euclidean $\vee$-systems only under additional
assumption that all the corresponding subspaces are non-isotropic.
In any case the complex Euclidean $\vee$-systems seem to be of independent interest and deserve further investigation.

\section{Concluding remarks}

We have seen that the $\vee$-systems have very interesting geometric
properties and some intriguing relations. The most important open
problem is their classification. It is open already in dimension 3.
In Figure 1 we pictured schematically all known non-reducible
$\vee$-systems in dimension 3.

\begin{figure}
\centerline{\includegraphics[width=24cm]{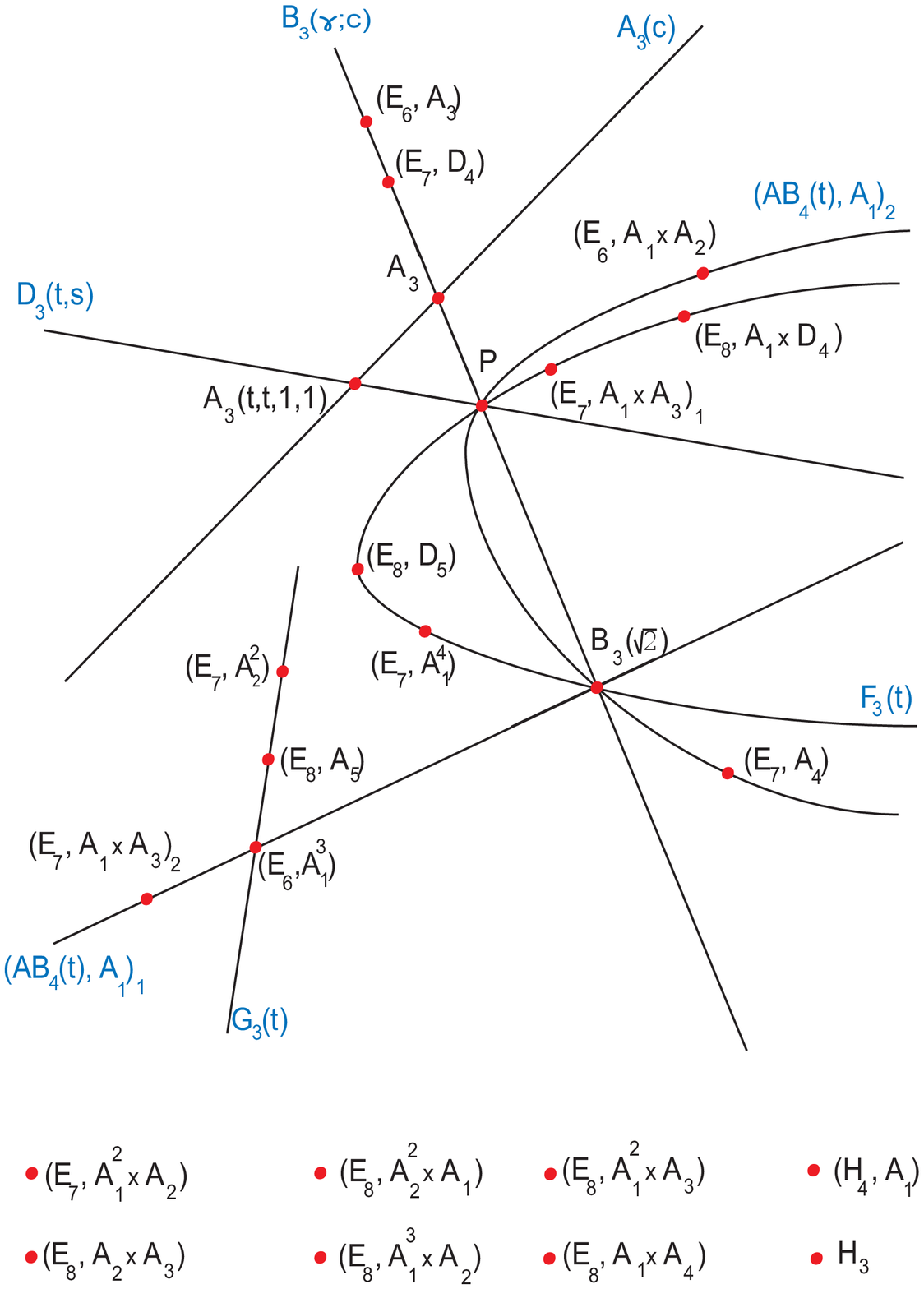}} \caption{All
known $\vee$-systems in dimension 3.}
\end{figure}

 All the curves in the diagram represent one-parameter  families of
$\vee$-systems except the curves corresponding to $A_3-$, $B_3-$ and
$D_3-$families. The families $A_3(c)$ and $B_3(\gamma;c)$
essentially depend on three parameters (after scalar dilatation of
all the vectors in a $\vee$-system) and $D_3(t,s)$ is a
two-parametric family of $\vee$-systems. The point $P$ on the
diagram represents the $\vee$-system $B_3(-1;1,1,2)$ which is the
intersection of the one-parameter families $(AB_4(t),A_1)_2$ and
$F_3(t)$. Also the point $P$ corresponds to the one-parameter family
$B_3(-1; 1,1, s)$ which is the intersection of $D_3-$ and $B_3-$
families.

In the diagram we used Theorems \ref{teorg}, \ref{teord} and
equivalences established in \cite{FV}. We also used that in the
limit $t \to \infty$ the restrictions of $AB_4(t)$-systems are
equivalent as follows:
$$
(AB_4(\infty), A_1)_1=B_3(\sqrt{2}),\quad (AB_4(\infty),
A_1)_2=B_3(-1; 1,1,2).
$$

{\bf Acknowledgements.} We are grateful to O.A. Chalykh and A.N. Sergeev for useful
discussions. The work was partially supported by the European
research network ENIGMA (contract MRTN-CT-2004-5652), ESF
programme MISGAM and EPSRC (grant EP/E004008/1).

We also acknowledge support of the Isaac Newton
Institute for Mathematical Sciences for the hospitality during September 2006,
when part of the work was done.

\end{document}